\documentclass[apj]{emulateapj}
\usepackage{epsf}
\newcommand{\Msun}{\mathrm{M}_{\odot}}

\newcommand{\trir}{\tilde{r}}
\newcommand{\mkas}{\mu\mathrm{as \, yr}^{-1}}
\newcommand{\mas}{\mathrm{mas \, yr}^{-1}}
\newcommand{\bdv}[1]{\mbox{\boldmath$#1$}}
\def\btheta{{\bdv{\theta}}}
\def\bv{{\bdv{v}}}
\def\be{{\bdv{e}}}
\def\bR{{\bdv{R}}}

\begin{document}

\shortauthors{GNEDIN ET AL.}
\shorttitle{SHAPE OF THE GALACTIC HALO}

\title{Probing the Shape of the Galactic Halo with Hyper-Velocity Stars}

\author{ Oleg Y. Gnedin\altaffilmark{1},
         Andrew Gould\altaffilmark{1},
         Jordi Miralda-Escud\'e\altaffilmark{1},
         Andrew R. Zentner\altaffilmark{2}
}

\begin{abstract}
Precise proper motion measurements ($\sigma_\mu \sim 10 \, \mkas$) of
the recently discovered hyper-velocity star (HVS) SDSS
J090745.0+024507 would yield significant constraints on the axis
ratios and orientation of a triaxial model for the Galactic halo.
Triaxiality of dark matter halos is predicted by Cold Dark Matter models
of galaxy formation and may be used to probe the nature of dark matter.
However, unless the distance to this star is determined to better than
10\%, these constraints suffer from one-dimensional degeneracies,
which we quantify.  We show how proper motion measurements of several
HVSs could simultaneously resolve the distance degeneracies of all
such stars and produce a detailed picture of the triaxial halo.
Additional HVSs may be found from radial velocity surveys or from
parallax/proper-motion data derived from GAIA.  High-precision
proper-motion measurements of these stars using the {\it Space
Interferometry Mission (SIM PlanetQuest)} would substantially tighten
the constraints they yield on the Galactic potential.
\end{abstract}


\altaffiltext{1}{Department of Astronomy,
       The Ohio State University, 
       140 W 18th Ave., Columbus, OH 43210;
       \mbox{\tt ognedin@astronomy.ohio-state.edu}}
\altaffiltext{2}{Kavli Institute for Cosmological Physics and
       Department of Astronomy and Astrophysics,
       The University of Chicago, Chicago, IL 60637}

\section{Introduction}

In the course of a spectroscopic survey of candidate faint blue
horizontal-branch stars found in the Sloan Digital Sky Survey,
\citet{brown_etal05} have discovered a star with the heliocentric
radial velocity $v_{\rm obs} = +853 \pm 12$ km s$^{-1}$ at the
Galactic coordinates $b = 31.3319^\circ$, $l = 227.3353^\circ$.  The
velocity of this star, SDSS J090745.0+024507 (hereafter called ``the
hyper-velocity star'', or HVS), is more than twice the escape speed
from the Galaxy at the star's present location.  Corrected for the
solar motion relative to the local standard of rest
\citep[LSR;][]{dehnen_binney98}, the line-of-sight (los) LSR velocity
is $v_{\rm los} = 848$ km s$^{-1}$.  The distance of the HVS from
Earth is presently uncertain due to an ambiguity in its spectral
classification.  If it is a main sequence B9 star, its heliocentric
distance is $d \approx 70$ kpc; if it is a blue horizontal branch
star, then $d \approx 40$ kpc.  A higher S/N spectrum from a large
ground-based telescope should determine the spectral type of this
star.

As discussed by \citet{brown_etal05}, the velocity of the HVS greatly
exceeds that plausible for a runaway star ejected from a binary in
which one component has undergone a supernova explosion.  The only
known mechanism for a star to obtain such an extreme velocity is
ejection from the deep potential of the massive black hole at the
Galactic center, as a result of scattering with another star or tidal
breakup of a binary \citep{hills88, yu_tremaine03}.  Only extremely
close to the massive black hole, at $r \lesssim 0.01$ pc, can stars
attain the required speeds $v \approx (2GM_{\rm bh}/r)^{1/2} \gtrsim
1000$ km s$^{-1}$.

A precise measurement of the three-dimensional motion of this HVS can
probe the shape of the mass distribution of the Galactic halo in a way
that is independent of any other technique attempted so far.  The
expected trajectory of the HVS in the Galaxy is almost a straight
line, owing to its extremely high velocity.  However, assuming its
origin at the Galactic center, the direction of the HVS's present
velocity should deviate slightly from being precisely radial due to
departures from spherical symmetry of the Galactic potential.

The main sources of such asymmetry are the flattened disk and the
(possibly) triaxial dark matter halo.  Several estimates of the halo
shape based on observations of tidal debris associated with the
Sagittarius dwarf galaxy indicate that it is close to spherical
\citep[e.g.][]{ibata_etal01,majewski_etal03,johnston_etal05}, although
\citet{helmi04a} argues that minor-to-major axis ratios as low as 0.6
for isodensity contours cannot be ruled out for a prolate halo
oriented with its major axis along the rotation axis of the disk.
While cosmological N-body simulations based on Cold Dark Matter models
typically produce prolate halos with density axis ratios in the range
$0.5-0.8$ \citep[e.g.][]{jing_suto02, bullock02}, gasdynamics
simulations indicate that the effects of gas cooling and dissipation
tend to make the halos rounder \citep{kazantzidis_etal04}.  Whether
the predictions of gasdynamics simulations agree with the observations
of tidal streams still remains to be seen.

Measuring the proper motions of the HVS and reconstructing its
three-dimensional velocity will provide useful constraints on the
shape and orientation of the Galactic dark matter halo, as we discuss
below.  Independent observational constraints of halo shapes are
important for testing structure formation models as well as probing
the nature of dark matter.

\section{Orbits of the Hyper-Velocity Star}

In order to evaluate the deviation of the HVS orbit from a straight
line, we have calculated a family of orbits of the HVS consistent with
its position in the sky (Galactic coordinates $l$ and $b$), the
assumed distance from Earth ($d$), and the observed line-of-sight
velocity, $v_{\rm los}$.  All orbits start at the origin $r=0$ with
some initial ejection velocity $\bv_i$, which is then adjusted until
the orbit reproduces the four observables.

We approximate the Galactic potential by the sum of three components:
spherical bulge, axisymmetric disk, and triaxial dark matter halo,
with parameters that are consistent with the current mass model of
the Galaxy by \citet{klypin_etal02}.  The bulge potential is given by
the \citet{hernquist90} model:
\begin{equation}
  \Phi_b(r) = -{G M_b \over r+a_b},
\end{equation}
with mass $M_b = 10^{10}\ \Msun$ and core radius $a_b = 0.6$ kpc.
The disk potential is given by the analytical \citet{miyamoto_nagai75}
model:
\begin{equation}
  \Phi_d(R,z) = -{G M_d \over
     \sqrt{ R^2 + \left({ a_d + (z^2 + b_d^2)^{1/2} }\right)^2 }}
\end{equation}
with mass $M_d = 4\times 10^{10}\ \Msun$, scale length $a_d = 5$
kpc, and scale height $b_d = 0.3$ kpc.  The halo potential is derived
from a generalized triaxial density distribution of the NFW model
\citep{NFW97}, which provides a good fit to halo profiles found in
cosmological N-body simulations \citep{jing_suto02}:
\begin{equation}
  \rho_h(\trir) = {M_h \over 4\pi \trir (\trir + r_s)^2},
  \label{eq:trinfw}
\end{equation}
with mass $M_h = 10^{12}\ \Msun$ and scale radius $r_s = 20$ kpc.
We generalize the halo profile as a function of the triaxial radius
$\trir$ to allow for three independent scale lengths:
\begin{equation}
  \trir^2 \equiv (x/q_1)^2 + (y/q_2)^2 + (z/q_3)^2.
\end{equation}

We explore the full range of the axis ratios consistent with current
observational constraints, $0.5 \le q_i \le 1$, to investigate the
maximum effect of halo triaxiality on the expected proper motions of
the HVS.  Specifically, we calculate three sets of models with the
halo major axis aligned with each of the three coordinate axes while
varying the other two axis ratios from 0.5 to 1 with a step 0.05.  For
simplicity we assume that the axis ratios are constant as a function
of radius, an assumption that is supported by the results of
cosmological $N$-body simulations \citep[e.g.][]{jing_suto02}.  In
practice, the motion of the HVS probes the halo shape in the range of
radii between 20 and 80 kpc (see Fig. \ref{fig:th_gc}).  We also do
not include the effect of adiabatic contraction of the halo in
response to the central concentration of baryons
\citep{blumenthal_etal86, ryden_gunn87, gnedin_etal04} since this
effect is likely to be small at large radii, $r \ga r_s$.

The triaxial halo potential is obtained by numerical integration over
thin triaxial shells, or homeoids \citep[e.g.][Chapter
2.3]{binney_tremaine87}.  Interior to the shell the potential is
constant, while isopotential surfaces exterior to the shell are
confocal ellipsoids labeled by the parameter $\tau$:
\begin{equation}
  \mathrm{const} = m^2 \equiv q_1^2 \sum_i {x_i^2 \over q_i^2 + \tau}.
\end{equation}
The exterior potential due to each such shell $m$ is proportional to
$\psi(\infty)-\psi(m[\tau])$, where
\begin{equation}
  \psi(m) \equiv \int^{m^2}_0 \rho(m^2) dm^2 
          = {M_h \over 2\pi r^2} {m \over m + r_s},
\end{equation}
and the second equality is for the density profile given by equation
(\ref{eq:trinfw}).  The total potential is the integral over all
shells,
\begin{eqnarray}
  \Phi_h(x,y,z) && = -{G M_h \over 2} {q_2 q_3 \over q_1} \nonumber\\
                \times \int_0^\infty && {d\tau \over (m + r_s)
                \sqrt{(q_1^2 + \tau)(q_2^2 + \tau)(q_3^2 + \tau)}}.
  \label{eq:triphi}
\end{eqnarray}
We have tabulated the integral in equation (\ref{eq:triphi}), as well
as the corresponding integrals for the force components, on a
three-dimensional grid of coordinates ($x,y,z$) and interpolated the
tables for orbit calculation.

With our definition of the triaxial profile (eq.~[\ref{eq:trinfw}])
the halo mass enclosed within a given spherical radius $R_{\rm vir}$
depends on the axis ratios.  The more triaxial halos have less
enclosed mass.  On the other hand, the measured mass of the Galaxy,
inferred from the radial velocities of distant satellites and globular
clusters, is in principle independent of halo triaxiality.  In order
to fix this discrepancy, we have renormalized the mass $M_h$ for
triaxial halo cases such that the mass within a spherical virial
radius $R_{\rm vir} = 12 \, r_s$ \citep[see][]{klypin_etal02} is the
same for all models and equals $10^{12}\ \Msun$.

\begin{figure}[t]
\centerline{\epsfysize3.5truein \epsffile{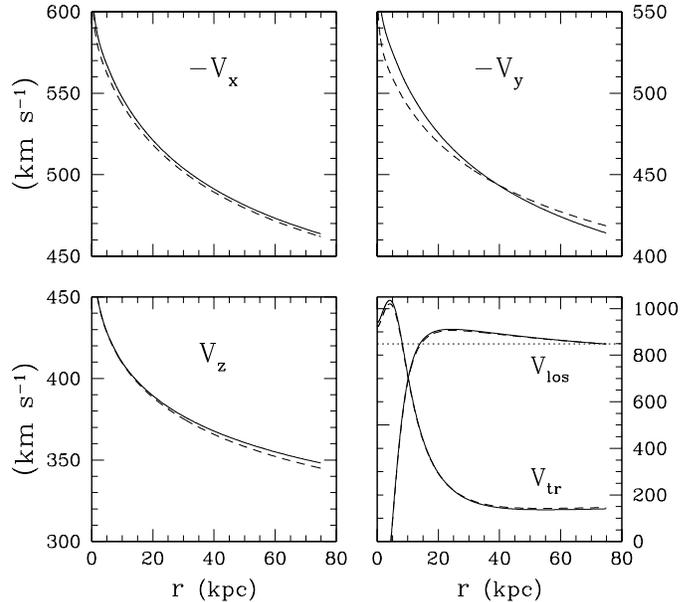}}
\caption{Deceleration of the HVS as a function of distance to the
  Galactic center.  Three velocity components in the Galactocentric
  frame are shown for the cases of a spherical halo (dashed lines) and
  a triaxial halo (solid lines).  The axis ratios of the triaxial
  model are $q_1/q_3 = 0.9$, $q_2/q_3 = 0.7$, and the assumed current
  distance from Earth is $d = 70$ kpc.  The bottom right panel shows
  the line-of-sight velocity as seen from Earth and the absolute value
  of the transverse velocity.  The observed los velocity is shown by a
  horizontal dotted line.
  \label{fig:v_vs_d}}
\end{figure}

\begin{figure}[t]
\centerline{\epsfysize3.5truein \epsffile{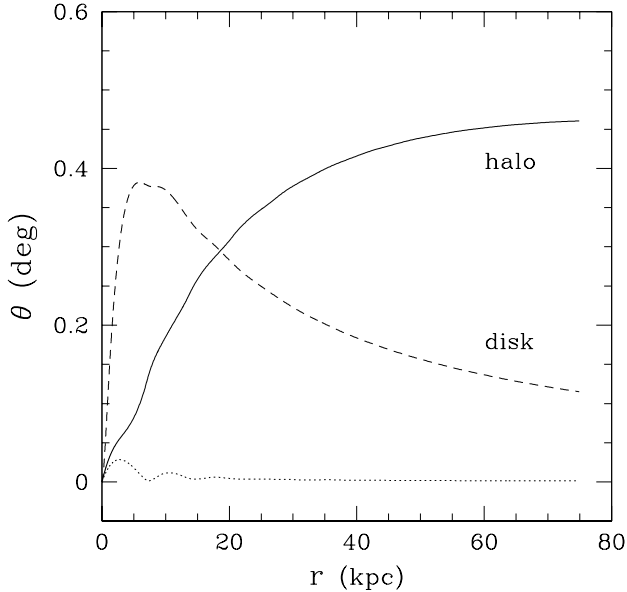}}
\caption{Angle between the radius vector and the velocity vector of
  the HVS in the Galactocentric frame vs the distance to the Galactic
  center, for the orbits shown on Fig. \protect\ref{fig:v_vs_d}.
  Solid line shows the apparent deflection of the orbit from a
  straight line due to the triaxial halo, without the
  flattened disk.  Dashed line shows the contribution of the disk
  alone, in the case of a spherical halo.  At large distances from the
  center, most of the deflection is due to the asphericity of  the
  halo.  As a check of numerical accuracy, dotted line shows that in a
  spherical bulge+halo potential without the disk, the inferred angle
  is close to zero as expected.
  \label{fig:th_gc}}
\end{figure}

Figure \ref{fig:v_vs_d} illustrates the deceleration of the HVS on its
way from the Galactic center for both a spherical halo model and a
triaxial halo model.  A typical ejection velocity is 900 km s$^{-1}$
in the Galactocentric frame, which is reduced to 700 km s$^{-1}$ at
the present position.  The line-of-sight direction varies rapidly at
small distances $r < 20$ kpc, when the HVS is relatively close to
Earth, which leads to large apparent variations of the transverse
velocity.  At larger distances, a significant part of $v_{\rm tr}$
represents the reflex motion of the Sun around the Galactic center.

For this and subsequent figures, we have chosen a fiducial triaxial
halo model with the axis ratios $q_1/q_3 = 0.9$, $q_2/q_3 = 0.7$, and
the major axis aligned with the Z-coordinate.  Our choice is motivated
by the orientation of satellite galaxies and their counterparts in
cosmological simulations \citep[e.g.,][]{zentner_etal05,
libeskind_etal05}, which indicates that the dark matter halo may be
prolate and oriented perpendicular to the plane of the disk.  The
prolate Galactic halo based on the kinematics of the Sagittarius dwarf
debris has been claimed by \citet{helmi04b} but disputed by
\citet{johnston_etal05}.  Even the evidence based on the satellites is
inconclusive at present \citep[c.f.,][]{navarro_etal04, bailin_etal05}
and our fiducial model should be considered as an example only.

Figure \ref{fig:th_gc} shows the deviation of the orbit from a
straight line, as measured by the angle between the radius vector and
the velocity vector in the Galactocentric frame.  Owing to the very
large ejection velocity from the center, this deviation is always
small, $\la 1\%$.  Therefore, the expected transverse velocities should
be of the order 1\% of the total velocity, or several km s$^{-1}$.

Figure \ref{fig:th_gc} also shows that the asymmetry of the potential
due to the flattened disk causes a smaller deflection than that due to
the triaxial halo, for stars at large distances.  In the case of a
spherical halo, the deflection contributed by the disk peaks at $r
\approx 10$ kpc but quickly declines at larger distances where the
disk density vanishes and the direction of the orbit aligns with the
velocity vector.  Because of this geometric effect, the angle
$\theta_{\rm disk}$ falls inversely proportional to the distance at $r
> 40$ kpc.  On the other hand, the halo density profile is still close
to isothermal at these distances, and the potential quadrupole induced
by halo triaxiality, $\delta\Phi \propto \rho_h(r) r^2$, is a weak
function of $r$.  This quadrupole causes a continuous deflection of
the orbit, $\delta v \sim \delta\Phi/v$.  Hence, measuring the
deviation of the trajectory of the HVS from a radial direction is
sensitive to the halo triaxiality and relatively insensitive to any
uncertainties in the mass model of the disk.

\begin{figure}[t]
\centerline{\epsfysize3.5truein \epsffile{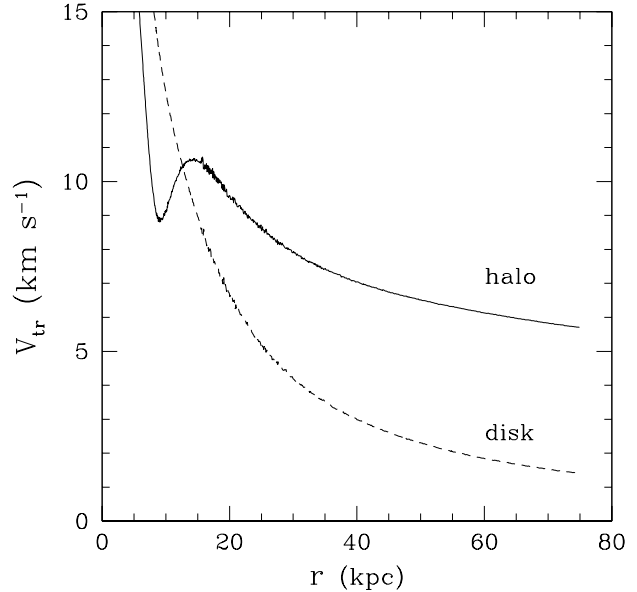}}
\caption{Difference between the velocities perpendicular to the
  line of sight (solid line) for the orbits in the triaxial and
  spherical halos shown on Fig. \protect\ref{fig:v_vs_d}.  For
  comparison, dashed line shows the difference between the transverse
  velocities in the cases of a spherical halo with and without the
  flattened disk.  These differences in $v_{tr}$ lead to the
  differences in the expected proper motions of the HVS for different
  shapes of the Galactic halo.
  \label{fig:v_tr}}
\end{figure}

A measurement of the proper motion of the HVS would give the
transverse velocity in the reference frame associated with Earth (or
after appropriate corrections, with the LSR).  In addition to the
perpendicular component of the Galactocentric velocity $\bv$, the
measured velocity would include the transverse component of the Solar
motion around the Galaxy, $\bv_{\odot}$, as well as a correction due
to the Earth being at a distance $R_\odot = 8$ kpc from the Galactic
center.  At large distances, $d \gg R_\odot$, where the angle $\theta
\ll 1$, the observed transverse velocity is
\begin{equation}
  \bv_{tr,\rm obs} \approx v \btheta 
      + v {\bR_\odot \times \hat{\be}_{\perp} \over d} 
      + \bv_{tr\odot},
  \label{eq:vtr}
\end{equation}
where $\btheta$ is the vector of length $\theta$ perpendicular to the
line of sight, $\bR_\odot$ is the vector of length $R_\odot$ towards
the Galactic center, and $\hat{\be}_{\perp}$ is the unit vector
perpendicular to both $\bR_\odot$ and the line of sight.  The last two
components depend only on the distance and angular coordinates of the
HVS but not on the shape of the halo and the disk.  For a given
position of the HVS, the differences in the expected proper motions
are only due to the deflections by the halo and the disk.

Figure \ref{fig:v_tr} shows the differences of the velocities
perpendicular to the line-of-sight for the triaxial and spherical
halos.  For the chosen halo model, $\Delta v_{tr} \approx 6$ km
s$^{-1}$.  This velocity difference is again dominated by the
triaxiality of the halo, as is the case with the deflection angles.
At $r > 10$ kpc, the contribution of the flattened disk decreases as
$\Delta v_{tr,\rm disk} \propto d^{-1}$.  The halo contribution also
decreases at large radii but only because of the smaller projection of
the Solar reflex motion along the orbit (third term in
eq.~[\ref{eq:vtr}]).  By construction of the orbits that reproduce the
observed positions of the HVS, at its current distance $\Delta v_{tr}$
is independent of the solar motion.

\begin{figure}[t]
\centerline{\epsfysize3.6truein \epsffile{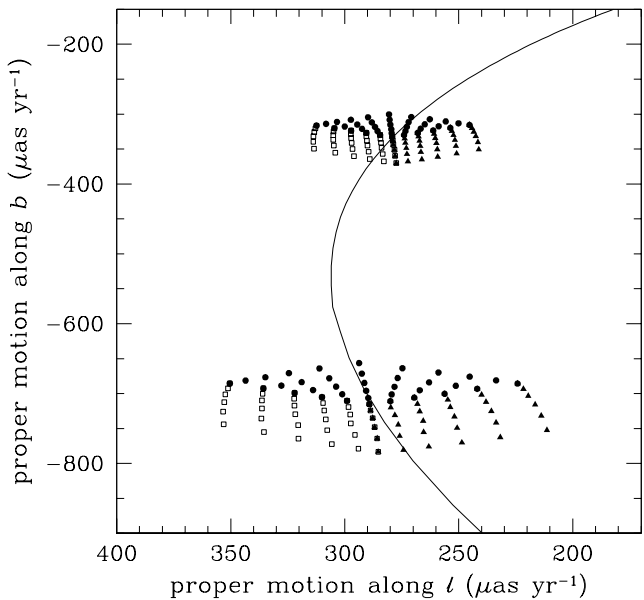}}
\caption{Expected proper motions of the HVS under a range of different
  assumptions about the shape and orientation of the Galactic dark
  matter halo.  The two components of the proper motion are in the
  directions of increasing Galactic coordinates, $l$ and $b$.  The
  family of models with the halo major axis along the Galactic
  X-coordinate is shown by triangles, along Y-coordinate by open
  squares, and along Z-coordinate by filled circles.  The upper and
  lower groups of the proper motions are for two assumed distances of
  the HVS of 70 kpc and 40 kpc, respectively.  The solid line shows
  the predicted proper motions for a HVS distance from Earth varying
  from 30 (bottom) to 140 kpc (top), assuming a spherical dark matter
  halo.
  \label{fig:prop}}
\end{figure}

\section{Reconstruction of the Halo Shape}

Placing constraints on the shape and orientation of the Galactic halo
is important for testing current models of galaxy formation.  Current
ground-based measurements of the proper motion of the HVS are
inconclusive and consistent with zero within large uncertainties
\citep[see][]{brown_etal05}.  However, the exquisite resolution of the
{\it Hubble Space Telescope (HST)} would allow one to measure the
proper motion of the HVS, or place useful limits on its value, within
just a few years.

The predicted proper motions in Galactic angular coordinates are
plotted in Figure \ref{fig:prop}.  The velocity transverse to the
line of sight is projected on two axes in the directions of
increasing Galactic coordinates $l$ and $b$.  The two groups of points
represent sets of triaxial halo models at two discrete distances,
while the solid curve represents a spherical model at a continuum of
distances.  The absolute values of the proper motions vary from $400
\, \mkas$ to $900 \, \mkas$, depending on which distance is chosen for
the HVS (70 and 40 kpc, respectively).

Note that the loci of the two distributions of the proper motions for
two assumed distances are well separated from each other, and
therefore the true distance of the HVS can be cleanly determined from
a proper-motion measurement accurate to $\sigma_\mu \sim 100 \,
\mkas$.

With an accuracy of $\sigma_\mu \sim 20 \, \mkas$ we can place
interesting constraints on the orientation of the Galactic halo.  In
particular, if the major axis of the halo is aligned with the
direction to the Galactic center ($l=0$), the predicted components
$\mu_l$ all lie below $\mu_l < 280 \, \mkas$.  If the major axis is in
the direction of solar rotation, $\mu_l > 300 \, \mkas$.  If the halo
is prolate with the major axis aligned with the disk rotation axis,
then the predicted $\mu_b$ component of the proper motion is well
constrained to be either $-320 \pm 20$ or $-700 \pm 40 \, \mkas$, for
the assumed distances of 70 and 40 kpc, respectively.  These estimates
are valid under our assumption that the halo axes are aligned with the
Galactic coordinate axes, i.e., $X$ along $l=0$, $Y$ along
$l=90^\circ$, and $Z$ perpendicular to the disk plane.

\begin{figure}[t]
\centerline{\epsfysize3.6truein \epsffile{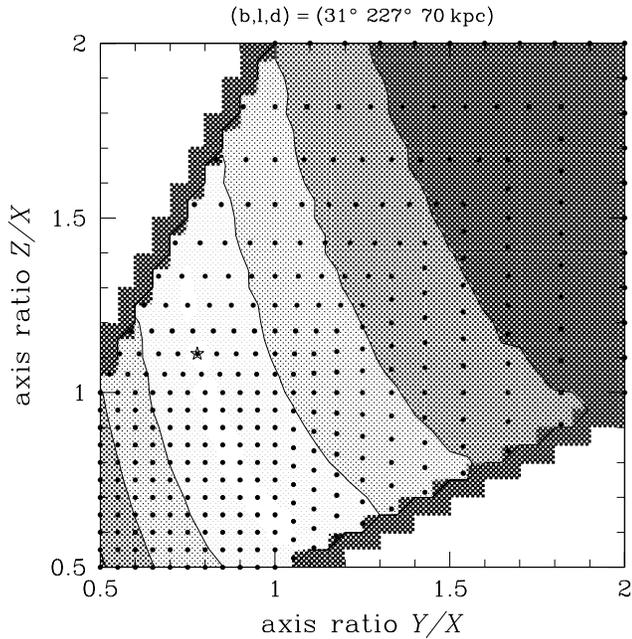}}
\caption{Reconstruction of the halo shape from the (assumed) measured
  proper motions of the HVS.  All calculated triaxial models are
  marked by filled dots corresponding to their axis ratios.  Lines
  represent 1$\sigma$, 2$\sigma$, and 3$\sigma$ contours, assuming
  measurement errors of $10 \, \mkas$.  The assumed proper motions
  correspond to the triaxial model with the axis ratios $q_1/q_3 =
  0.9$, $q_2/q_3 = 0.7$, which are marked by a star, and the distance
  of the HVS of $d=70$ kpc.
  \label{fig:obs}}
\end{figure}

With an accuracy of $\sigma_\mu = 10 \, \mkas$ we can attempt a
reconstruction of the halo axis ratios.  Figure \ref{fig:obs} shows an
example of such a future reconstruction given a hypothetical
measurement of the proper motions of the HVS, $\mu_l^{\rm obs}$ and
$\mu_b^{\rm obs}$.  The three sets of models (each with the major axis
aligned with one of the coordinate axes) are combined in a single plot
of the axis ratios $q_2/q_1 = Y/X$ versus $q_3/q_1 = Z/X$.  The lower
left part of the plot corresponds to the case of the major axis
along the Galactic X-coordinate ($q_1=1$), whereas the lower right and
upper left parts correspond to the major axis along the Y-coordinate
($q_2=1$) and the Z-coordinate ($q_3=1$), respectively.  In this plot
we calculate
\begin{equation}
  \chi^2 = {(\mu_l - \mu_l^{\rm obs})^2 + (\mu_b - \mu_b^{\rm obs})^2
           \over \sigma_\mu^2}
           + {(d - d_{\rm obs})^2 \over \sigma_d^2}
\end{equation}
assuming that the true distance of the HVS is unknown ($\sigma_d
\rightarrow \infty$) and search for all predicted proper motions
$\mu_l,\mu_b$ at all possible distances between 20 kpc and 100 kpc
that minimize $\chi^2$.  The assumed observed proper motions, however,
correspond to the fiducial model with $q_1/q_3 = 0.9$, $q_2/q_3 =
0.7$, and the distance of 70 kpc.

According to Figure \ref{fig:obs}, a certain range of the axis ratios
can be excluded based on the proper motion accuracy of $\sigma_\mu =
10 \, \mkas$.  However, there are strong triaxiality-distance
degeneracies.  The best-fitting axis ratios form a band stretching
across the diagram because different halo shapes with a somewhat
different target distance can produce similar proper motions.  This
effect is evident in Figure \ref{fig:prop}.  The inferred distances
for our fiducial model range from 67 to 78 kpc.  The allowed contours
would be reduced if the actual distance of the HVS were known, but it
needs to be measured to better than 10\% to make a significant
difference.  Also, if the true distance of the HVS is smaller than the
70 kpc assumed in our fiducial model, the constraints become tighter.

\begin{figure}[t]
\centerline{\epsfysize3.6truein \epsffile{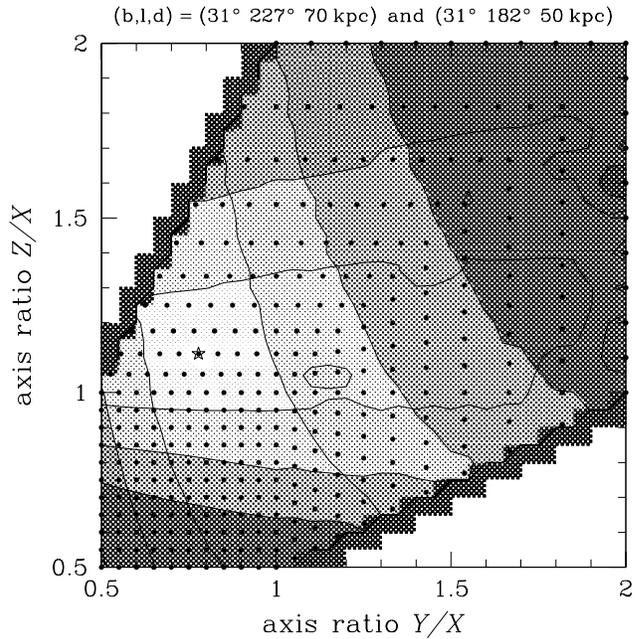}}
\caption{Combined reconstructed contours for two HVS, assuming another
  star were detected at a $45^\circ$ offset in the Galactic $l$ coordinate,
  at a distance of 50 kpc from Earth.
  \label{fig:31_182_50}
}
\end{figure}

\section{Two, Three, ... Many Hyper Velocity Stars}

\subsection{Two Hyper Velocity Stars}

Measurements of the proper motions of the single known HVS would
already provide interesting constraints on the shape and orientation
of the Galactic halo.  Yet, the allowed range of axis ratios is fairly
broad, even when the measurement errors are as small as $10 \, \mkas$.
The constraints could become tighter if another HVS were discovered.

For example, suppose that another hypervelocity star, HVS2, were
discovered at the same angle $b_2 = 31^\circ$ above the Galactic plane
but offset by $45^\circ$ in longitude, $l_2 = l - 45^\circ =
182^\circ$, and at a hypothetical distance of 50 kpc.  Figure
\ref{fig:31_182_50} shows the superposition of the contours of the
original HVS and the new HVS2.  The $\chi^2$ contours for the two
stars intersect almost perpendicularly to each other and significantly
reduce the acceptable range of halo shapes.  At the $1\sigma$ level the
axis ratios would be constrained to better than $20\%$.

\begin{figure}[t]
\centerline{\epsfysize3.6truein \epsffile{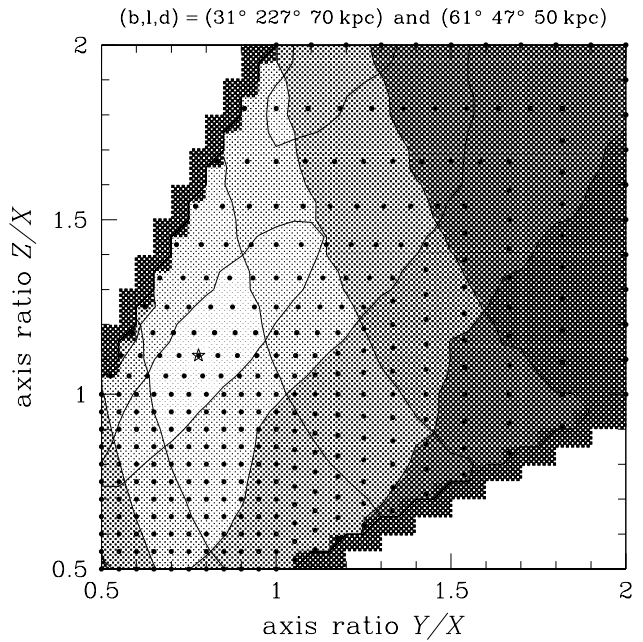}}
\caption{Same as Fig. \ref{fig:31_182_50}, but assuming an offset of
  $180^\circ$ in $l$ and $30^\circ$ in $b$.
  \label{fig:61_47_50}
}
\end{figure}

Do all directions of the HVS2 on the sky provide equally useful
constraints on the halo axis ratios?  Figure \ref{fig:61_47_50} shows
the contours for the HVS2 offset by $180^\circ$ in longitude ($l_2 =
47^\circ$) and $30^\circ$ in latitude ($b_2 = 61^\circ$).  These
contours also intersect the contours for the original HVS and the
resulting constraints are also tight.  However, in the case (not
shown) that the two stars were at the same longitude ($l_2 =
227^\circ$) but only offset in latitude ($b_2 = 61^\circ$), the
contours of the HVS2 are almost parallel to those of the HVS and
provide no new information.  Thus, searching prospective HVSs at
different longitudes than the original HVS would yield the most
interesting new constraints.

\subsection{Three or more Hyper Velocity Stars}

Although there are physical motivations (albeit not conclusive) for
the halo to be prolate with its major axis perpendicular to the plane
of the Galactic disk \citep[e.g.][]{helmi04a,zentner_etal05}, the
orientation of the other two axes in the plane of the disk is somewhat
arbitrary.  There is nothing special about the current location of the
Sun, which defines the Galactic $X$-coordinate.  Therefore, the halo
axes in the plane can be rotated with respect to the Galactic
coordinates by any angle between 0 and $90^\circ$.  This means that
even under the most optimistic assumptions, a minimum of three HVSs
are required to fix the Galactic potential.

\begin{figure}[t]
\centerline{\epsfysize3.6truein \epsffile{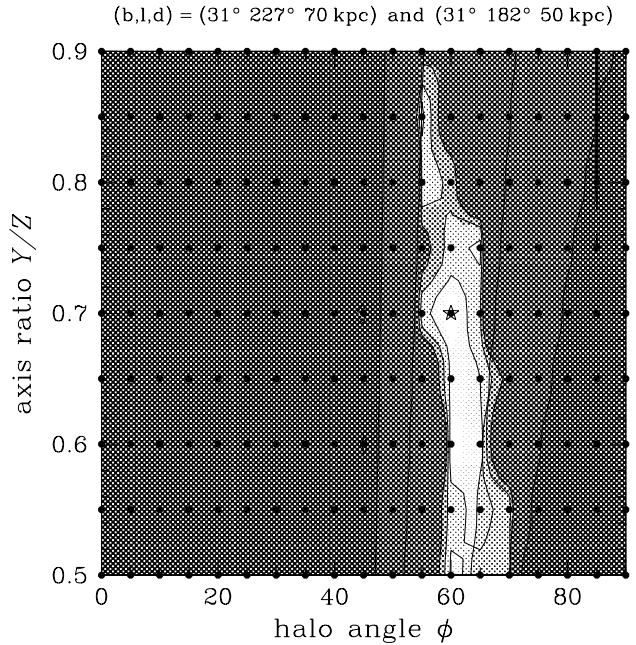}}
\caption{Same as Fig. \ref{fig:31_182_50}, but as a function of angle
  between the halo axis $q_1$ and the Galactic $X$-coordinate.  The
  axis ratio $q_1/q_3 = 0.9$ is kept fixed in this plot.
  Assumed measurement errors are $30 \, \mkas$.
  \label{fig:obs_angle}}
\end{figure}

To illustrate the role of three HVSs, we have calculated another set
of models, keeping fixed the ratio $q_1/q_3 = 0.9$ and varying two
parameters: the ratio $q_2/q_3$ between 0.5 and 0.9, and the angle
$\phi$ between the halo axis $q_1$ and the Galactic $X$-coordinate
between $5^\circ$ and $90^\circ$.  Figure \ref{fig:obs_angle} shows
the resulting contours for a fiducial model in which the $q_1$ axis is
rotated by $60^\circ$.  The rotation angle is constrained very well,
within $5^\circ-10^\circ$, even if we assume larger measurement
errors, $\sigma_\mu = 30 \, \mkas$.

We have also explored the configuration (not shown) in which the halo
is oblate rather than prolate, with the minor axis along the rotation
axis of the disk.  In this configuration the angular momentum vectors
of the halo and the disk are parallel to each other, as has often been
assumed in the literature.  We find that the contours in this model
are qualitatively similar to those in Fig. \ref{fig:obs_angle}, and in
particular, angle $\phi$ can be constrained with a similar accuracy.

If five HVSs were detectable, it would be possible to measure all five
parameters of a more general triaxial halo.  These would be the two
axis ratios, and the three direction angles.  Additional HVSs (beyond
five) could be used either to improve the precision of the estimates
of the first five parameters or to explore halo models of additional
complexity.

\section{Discussion}

We have shown that proper motion measurements of the recently
discovered HVS, accurate to $\sigma_\mu \sim 10 \, \mkas$, would place
significant constraints on the axis ratios and orientation of the
Galactic halo.  With only one HVS, such constraints are limited by the
triaxiality-distance degeneracies.  They can be improved only when the
distance is determined to within a few kpc.  However, the discoveries
and proper motion measurements of several additional HVSs could
simultaneously resolve the distance degeneracies of all such stars and
produce tighter limits on the triaxiality of the Galactic halo.  We
now discuss implications of the discovery of multiple HVSs and
potential methods of finding more HVSs in future surveys.

\subsection{History of Ejection from the Galactic Center}

If several hyper velocity stars are found, it may be possible to
reconstruct the history of their ejections from the Galactic center
and therefore, constrain the rate of binary disruption near the
massive black hole.  \citet{yu_tremaine03} estimate the rate of
ejection from the breakup of binary stars of $\sim 10^{-5}$ yr$^{-1}$.
If there is another massive black hole in the close vicinity of the
central black hole at the Galactic center, then stars can be ejected
by three-body interactions with the binary system formed by the two
black holes.  \citet{yu_tremaine03} estimate that the rate of such
events $\la 10^{-4}$ yr$^{-1}$.  The density of the HVSs in the halo
can be calculated if HVSs are ejected from the Galactic center
isotropically at a steady rate $\dot{N} = 10^{-5} \, \dot{N}_{-5}$
yr$^{-1}$.  The expected number density in this case is $n(r) \approx
\dot{N}_{-5} \, r^{-2}$ kpc$^{-3}$, where $r$ is in kpc.

There are several circumstances that could lead to clumping of HVS
ejections in time, e.g., if bursts of star formation (lasting $\sim 1$
Myr) or infall of clusters containing intermediate-mass black holes
\citep[e.g.,][]{hansen_milosavljevic03} led to a rapid increase in the
ejection rate.  If the global fit to multi-HVS data led to distance
estimates accurate to 5 kpc, it would enable one to date individual
ejections with an accuracy of about 5 Myr, and therefore, to test
whether the ejections were clumped in time.

\subsection{How to find HVS?}

There are basically two potential methods to find more HVSs: proper
motions and radial velocities.  For practical reasons, these methods
probe two different volumes, the ($\sim 10$ kpc) ``solar
neighborhood'' and the outer Galaxy, respectively.

If a star ejected from the Galactic center is passing within $\sim 10$
kpc of the Sun with a speed significantly greater than the escape
speed, then its proper motion will be of order $\mu \sim 1000 \, \rm
km\,s^{-1}/10\,kpc \sim 20 \, \mas$.  Given a proper motion survey
with precision $\sim 4 \, \mas$, this would be detectable at the $5 \,
\sigma$ level.  Such precision is already available for the several
thousand square degrees covered by the Sloan Digital Sky Survey
\citep{gould_kollmeier04, munn_etal04}.  However, robust
identification of high-velocity candidates requires not just reliable
proper motions, but reliable distances as well.  While horizontal
branch (HB) stars are approximately standard candles whose distances
can be reliably estimated from their colors and magnitudes, halo blue
stragglers (BS) can masquerade as HB stars.  A BS that was 2 mag
dimmer than a HB star would have the same proper motion as a HB HVS
but would actually be travelling 2.5 times slower.  That is, an
ordinary halo BS could easily be misinterpreted as a HB HVS.
Similarly, because subdwarfs are often 2 mag dimmer than main-sequence
stars of the same color, ordinary subdwarfs could be mistaken for HVS
main-sequence stars.  To some extent, HVS candidates can be singled
out because their trajectories ``point back'' to the Galactic center.
However, since halo stars tend to be on radial orbits, this
characteristic is not so distinguishing unless the proper motions are
fairly precise.

The situation will change radically when GAIA data become available,
not only because of its $4\pi$ sky coverage but also because it will
provide parallaxes as well as proper motions.  At present, these are
expected to be $\sigma_\pi \sim 20 \, \mu\mathrm{as}$ at $V=15.5$,
implying roughly 20\% parallax errors for HB stars at 10 kpc.  Dimmer
stars would be similarly measurable but only within a much smaller
volume.

The proper-motion technique does not work at all for HVS stars like
SDSS J090745.0+024507 that have already left the Galaxy.  The
transverse velocities of these stars in the Galactic frame are only
$v_{tr} \sim v (R_\odot/d)$, where $v$ is their actual velocity and
$d$ is their distance (see eq. [\ref{eq:vtr}]).  This is similar to
the value for halo stars, making them not easily distinguishable.
Rather, these distant HVS can be reliably identified using
radial-velocity surveys of distant stars (as was done for SDSS
J090745.0+024507).  The only drawback is that many stars with large
photometric distances must be measured to find a small number of HVSs.
Since the number density of HVS falls as $r^{-2}$, while the density
of contaminating Galactic stars falls as $r^{-3.5}$, the contamination
rate falls for more distant samples.  However, more distant stars
require longer integration times.  Moreover, once found, they require
more accurate proper motions (on fainter stars!) to derive the same
precision on the transverse velocity, and so to extract similar
information about the Galactic potential.  Thus, there do not appear
to be any easy paths to finding HVSs.

Finally, we remark that the simulations we have conducted have assumed
proper-motion errors of $\sigma_\mu \sim 10-50 \, \mkas$, the upper
end of which is roughly what is achievable using the {\it HST}
Advanced Camera for Surveys with a 4-year baseline.  It is open to
question whether {\it HST} will even be operational in 4 years, let
alone in the additional time required to first find the HVSs.  Over
the longer term, the {\it Space Interferometry Mission (SIM
PlanetQuest)} expects to achieve $\sigma_\mu \sim 4 \, \mkas$ for
targeted stars with $V \la 20$.  For stars toward this faint limit,
this very high precision can be achieved only by quite long
integrations, typically several tens of hours.  However, the
integration time falls inversely as the square of the desired
precision, so that exposure times could be fine-tuned to the precision
required for each specific HVS.

\acknowledgements 
We acknowledge discussions at the OSU Astro Coffee that initiated this
project.  OYG acknowledges support from NASA ATP grant NNG04GK68G.
Work by AG is supported by grant AST 02-01266 from the NSF and by JPL
contract 1226901.  ARZ is supported by the Kavli Institute for
Cosmological Physics at The University of Chicago and the NSF under
grant PHY 0114422.

\bibliography{hvs}

\end{document}